\documentclass{PoS}

\title{Nucleon sigma term and 
strange quark content in 2+1-flavor QCD
with dynamical overlap fermions}

\ShortTitle{Nucleon sigma term and strange quark content in 2+1-flavor
with overlap quarks}

\author{\speaker{H. Ohki}$^{a,b}$, 
	S. Aoki$^{c,d}$
 	H. Fukaya$^e$, 
        S. Hashimoto$^{f,g}$, 
	T. Kaneko$^{f,g}$
        H. Matsufuru$^f$,\quad\quad\quad 
	J. Noaki$^f$, 
        T. Onogi$^h$, 
        E. Shintani$^h$, 
	N. Yamada$^{f,g}$ 
        (for JLQCD collaboration) \\
        \llap{$^a$}Department of Physics, 
         Kyoto University, Kyoto 606-8501, Japan,\\
        \llap{$^b$} Yukawa Institute for Theoretical Physics, 
        Kyoto University, Kyoto 606-8502, Japan, \\
	\llap{$^c$} High Energy Accelerator Research Center, 
	Brookhaven National Laboratory, Upton, New York 11973, USA \\
 	\llap{$^d$} School of High Energy Accelerator Science, The 
	Graduate University for Advanced studies(Sokendai), Ibaraki
 	305-0801, Japan \\
        \llap{$^e$} Department of Physics, Nagoya University, Nagoya
 	464-8602, Japan \\
        \llap{$^f$}
        High Energy Accelerator Research Organization (KEK), 
         Tsukuba 305-0801, Japan, \\
        \llap{$^g$}
        School of High Energy Accelerator Science, 
        The Graduate University for Advanced Studies (Sokendai), 
        Tsukuba 305-0801, Japan, \\
	\llap{$^h$} Department Physics, Osaka University, Toyonaka,
 	Osaka 560-0043, Japan \\
	E-mail: \email{ohki@yukawa.kyoto-u.ac.jp}}

\abstract{
We study the sigma term and the strange quark content of nucleon in 
2+1-flavor QCD with dynamical overlap fermions. We
analyze the lattice data of nucleon mass taken at two different
strange quark masses with five values of up and down quark masses
each. Using the reweighting technique, we study the strange quark 
mass dependence of the nucleon and extract the strange quark content.}

\FullConference{The XXVII International Symposium on Lattice Field Theory - LAT2009\\
		 July 26-31 2009\\
		 Peking University, Beijing, China}

\begin{document}

\section{Introduction}
Nucleon sigma term $\sigma_{\pi N}$ 
is given by a scalar form factor of nucleon at zero recoil. 
While up and down quarks contribute to $\sigma_{\pi N}$ both as
valence and sea quarks, 
strange quark appears only as a sea quark
contribution. 
As a measure of the strange quark content of the nucleon, 
the $y$ or $f_{T_s}$ parameters are commonly introduced.
These parameters are defined as 
\begin{equation}  \label{eq:piNsigma}
  \sigma_{\pi N} =  m_{ud} 
  \langle N | \bar{u}u+\bar{d}d | N \rangle, 
\quad \quad 
y  \equiv  
  \frac{2\langle N | \bar{s}s  | N \rangle}
  {\langle N | \bar{u}u +\bar{d}d | N \rangle},
\quad \quad
f_{T_s}=\frac{m_s \langle N |\bar{s}s  |N \rangle}{M_N}.
\end{equation}
The $y$ parameter plays an important role 
to determine the detection rate
of possible neutralino dark matter
 in the supersymmetric extension 
of the Standard Model  
\cite{Griest:1988yr,Ellis:2008hf}.

Using lattice QCD, one can calculate the nucleon sigma term 
directly.
Furthermore, it is possible to determine the valence and sea quark
contributions separately.
Previous lattice studies of the nucleon sigma term and 
strange quark content have been done 
within the quenched approximation
\cite{Fukugita:1995ba,Dong:1996ec},
or a two-flavor QCD calculation \cite{Gusken:1998wy,Ohki:2008ff}.
Recent studies use 
2+1 flavor QCD~\cite{Young:2009zb}. 

There was an apparent puzzle in these results: 
the strange quark content 
is unnaturally large compared to 
the up and down contributions 
that contain the connected diagrams too.
Concerning this problem, 
it was pointed out 
that using the Wilson-type fermions, 
the sea quark mass dependence of the additive mass renormalization 
and lattice spacing can give rise to 
a significant lattice artifacts
in the sea quark content~\cite{Michael:2001bv}.
Our previous study from a two-flavor QCD simulation 
using the overlap fermion 
removed this problem by explicitly maintaining exact chiral
symmetry on the lattice for the first time. It was revealed that the sea quark 
gives only a small contribution to the nucleon sigma term.

In this study, 
we analyze the data of the nucleon mass obtained 
from a 2+1-flavor QCD simulation 
employing the overlap fermion~\cite{Matsufuru}.
We employ the chiral perturbation theory (ChPT) analysis 
for the nucleon mass data and calculate the sigma term 
from its up and down quark mass dependence.
To calculate the strange quark mass dependence, 
we use the reweighting technique for the strange quark mass. 
Our study with exact chiral symmetry
avoids the contamination due to a significant lattice artifact 
so that it provides a reliable calculation of
the strange quark content.

\section{Lattice simulations}
\begin{figure}[tbp]
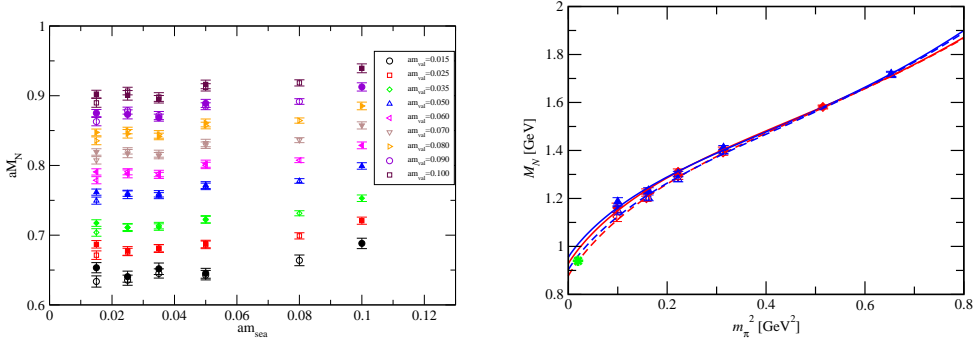

  \centering
  \rotatebox{0}{
    \includegraphics[width=6cm,clip]{./figure/meff_sea.eps}
\quad \quad 
    \includegraphics[width=6cm,clip]{./figure/nucl.eps}
  }
  \caption{
(Left) 
The nucleon mass data for the partially quenched data points.
The empty and full symbols show the fit results for $am_s=0.080$
and $am_s=0.100$, respectively.
(Right)
Chiral fit of the lattice raw data(full symbols) 
and the finite size corrected data(empty symbols).
The solid and dashed curve represent the fits of 
raw and corrected data.
For a reference, 
we also show the experimental values of the nucleon mass (star).}
\label{fig:fit}
\end{figure}

We make an analysis of the nucleon mass obtained on 
2+1-flavor QCD configurations generated with dynamical overlap
fermions. 
Our simulations are performed on a $16^3\times 48$ lattice with a trivial 
topological sector $Q=0$. 
For the gluon part, 
the Iwasaki action is used at $\beta=2.30$ 
together with 
unphysical heavy Wilson fermions and associated twisted-mass ghosts
\cite{Fukaya:2006vs} introduced to suppress unphysical near-zero modes
of $H_W(-m_0)$. 
The lattice spacing $a=0.108$ fm ($a^{-1}=1.83$ GeV) 
is determined through the Sommer scale $r_0=0.49$ fm of the static
quark potential.
We take two different strange sea quark masses $am_s=0.080$ and  $0.100$ 
with five values of degenerate up and down quark masses
(for $am_s=0.080$, $am_{sea}=0.015, 0.025, 0.035, 0.050, 0.080$
and for $am_s=0.100$, $am_{sea}=0.015, 0.025, 0.035, 0.050, 0.100$).
We accumulate about 2,500 
trajectories for each combination of sea quark masses;
the calculation of the nucleon mass is carried out at every 
five trajectories, thus we have 500 samples for each 
sea quark masses.
In order to improve the statistical accuracy, we 
use the low-mode averaging technique.
The two-point function made of the low-lying modes of the Dirac
operator is averaged over different times slices with 80 chiral pairs 
of the low modes.
For the quark propagator, we take a smeared-local two-point correlator
and fit the data with a single exponential function after 
averaging over forward and backward propagating states in time.
The statistical error is estimated using the starndard jackknife
method with a bin size of 10 samples.
We take the valence quark masses 
$am_{val}=$
0.015, 0.025, 0.035, 0.050, 0.060, 0.070, 0.080, 0.090 and 0.100.
All the plots of the nucleon mass spectrums are shown in the left
panel of Fig.~\ref{fig:fit}.

The matrix element defining the nucleon sigma term 
can be related to the quark mass dependence of the nucleon mass using
the Feynman-Hellman theorem.
For the strange sea quark, the matrix element is represented by the 
following relation as 
$  \frac{\partial M_N}{\partial m_{\rm s}} \
  = \langle N | \bar{s}s | N\rangle_{\rm disc}$.
The subscript ``disc'' on the expectation value
indicates that
only the disconnected quark line contractions are
evaluated. 
In the present study we exploit this indirect method 
to extract the matrix elements 
corresponding to the nucleon sigma term.

Another possible method is to directly calculate the
matrix element from three-point functions with 
an insertion of the scalar density operator 
$(\bar{u}u+\bar{d}d)(x)$ or $\bar{s}s(x)$.
We note that the indirect method gives  mathematically 
equivalent results to those by the direct method 
including lattice artifacts.
Numerical differences could arise only
in the statistical error and the systematic uncertainties 
due to the fit ansatz. 
\section{Analysis of the nucleon mass using
 chiral perturbation theory}\label{sec:sigma}

We carry out a fit with the SU(2) NLO ChPT for the baryon,
\begin{equation}
  \label{eq:p3}
  M_N = M_0 -4c_1 m_\pi^2 -\frac{3g_A^2}{32\pi f_\pi^2}m_\pi^3 
  +  e_1^r(\mu)  m_\pi^4,
\end{equation}
where $M_0$ is the nucleon mass in the chiral limit and
$f_\pi$ is the pion decay constant fixed 
at its physical value 92.4~MeV and 
the constant $g_A$ describes the nucleon axial-vector coupling.
We fix $g_A$ to its experimental value $g_A=1.267$.
We use the pion mass obtained from the calculation of the meson spectra.  
Since our lattice volume is not large enough, 
the finite size effects must be taken into account 
especially in the data at lightest quark mass. 
To estimate possible systematic error, we attempt the fit with and 
without lightest sea quark mass~(4 points and 5 points, respectively). 
We also correct the 
finite volume effect based on the NLO ChPT formula. 
The nucleon mass difference in the finite volume 
with box size $L$ and infinite volume
$\delta M_N(L)$ is calculated within the ChPT. 
We use the asymptotic formula~\cite{Beane:2004tw} 
to evaluate the magnitude of the finite 
volume effect as
\begin{equation}  \label{eq:fvc}
  \delta M_N(L) =\left( \frac{9g_A^2 m_\pi^2}{16\pi f_\pi^2}
  +\frac{4g_{\Delta N}^2 m_\pi^{\frac{5}{2}}}
{(2\pi)^\frac{3}{2}\Delta L^\frac{1}{2}} \right)
\frac{1}{L}\exp{(-m_\pi L)},
\end{equation}
where $\Delta$ is the $\Delta$-nucleon mass splitting,
and the coupling $g_{\Delta N}$ is decuplet-octet-axial coupling.
We use the phenomenological estimates 
$g_{\Delta N}=1.5$ and $\Delta=300$ MeV. 

The chiral fit is made for the corrected five data points
using the formula~(\ref{eq:p3}). 
The results are shown in the right panel of Fig.~\ref{fig:fit}. 
After correcting the finite volume effect, there are 6\% decrease in $M_0$
and 5\% increase in the magnitude of the slope $|c_1|$.
Finally we obtain the nucleon sigma terms by estimating the pion mass
dependence of the nucleon mass at the physical pion mass. 
The results are shown in Table~\ref{tab:sigma_FSE}. 
We observe that both the fit results at $am_s=0.080$ and $am_s=0.100$ 
without lightest points are consistent each other.
These results are also consistent with our previous results 
for two-flavor simulations.
We quote our best value of the sigma term as 
a fit result of 4 points fit ($am_s=0.080$, FVCs not included). 
We obtain $\sigma_{\pi N}=50.0(4.5)$. 
The finite volume correction of the sigma term amounts to about
$6\%$.
We note that a sizable chiral extrapolation error should also be 
added after comparing NLO and NNLO in ChPT as found in our 
previous analysis with two-flavor QCD.

\begin{table}[tbp]
  \centering
  \begin{tabular}{cccc}
    \hline
    &  \multicolumn{2}{c}{without FVCs} &  with FVCs \\
    & 4 points  & 5 points & 5 points \\
    \hline
    $am_s=0.080$   & 50.0(4.5) & 48.2(3.0)  & 51.1(3.0)  \\
    $am_s=0.100$   & 51.3(4.1) & 45.5(3.0)  & 48.2(3.0)  \\
   \hline
  \end{tabular}
  \caption{
    Nucleon sigma term $\sigma_{\pi N}$ [MeV]
    with and without the finite volume corrections~(FVCs).
  The error is statistical only.  } 
  \label{tab:sigma_FSE}
\end{table}

\section{Reweighting the strange quark mass} 

In order to study the strange quark mass dependence of the nucleon, 
we utilize the reweighting technique~\cite{Hasenfratz:2008fg,DeGrand:2008ps}.
Using this method we can calculate 
any physical observables at a slightly different strange quark mass
($m_s'$) from the configurations
($U$) generated at a certain value of strange quark mass ($m_s$) as 
\begin{equation}
  \label{eq:rew}
\langle \mathcal{O} \rangle_{m_s'}=
\frac{\langle \mathcal{O}(U,D(m_s')) w(m_s',m_s) \rangle }
{\langle w(m_s',m_s) \rangle}
\end{equation}
where $D$ is overlap Dirac operator and 
we introduce the reweighting factor $w(m_s',m_s)$ defined as
$w(m_s',m_s)=\det\left[ \frac{D(m_s')}{D(m_s)} \right]$.
We calculate the reweighting factor
by decomposing it into the low and high-mode 
contributions.
A former contribution is obtained exactly
by the 80 pairs of low-lying eigenvalues of the Dirac operator.
For high-mode, we use the improved stochastic estimator 
\begin{eqnarray}  \label{eq:high}
w_{\rm{high}}(m_s',m_s) =
\sqrt{
\frac{1}{N_r}
\sum_{r=1}^{N_r} e^{-\frac{1}{2}(\bar{P}\xi_r)^\dagger (\Omega-1)\bar{P}\xi_r}
}, \\
\Omega \equiv D(m_s)^\dagger \{D(m_s')^{-1}\}^\dagger D(m_s')^{-1}D(m_s),
\nonumber
\end{eqnarray}
where $\xi_r$ are the ensembles of pseudo-fermion fields
($N_r$: number of noise vector), and each are generated with Gaussian probability.
Here we also introduce the projection operators $P$ and $\bar{P}$ 
to project the eigenspaces of $\xi$ into low and high-modes, respectively. 

\begin{figure}[tbp]
  \centering
    \includegraphics[width=6cm,clip]{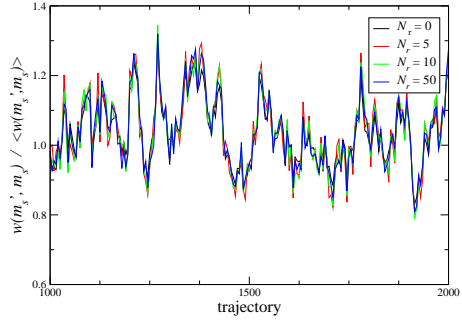}
\caption{
The fluctuations of the reweighting factor.
The plot shows the noise number dependence of the fluctuations 
at $am_{val}=am_{sea}=0.050$ with strange quark masses $am_s'=0.075$, $am_s=0.080$.
}
\label{fig:rwfctr}
\end{figure}

Fig.~\ref{fig:rwfctr}
shows the reweighting factor with $am_s'=0.075$ and $am_s=0.080$ 
for each gauge configuration.
It is evaluated with different number of noise vector 
$N_r=$0, 5, 10 and 50.
We find that the statistical error of the reweighting factor is very
 small compared to that from the fluctuation of the gauge configurations.
We also find that the contribution from high-mode is almost
independent of the gauge configuration and the configuration
dependence is almost completely dictated by the low-mode contribution.
These properties are universal for entire region of up and down quark
masses.

In order to estimate the strange quark mass dependence of the nucleon
mass, we evaluate the nucleon masses at 10 different strange quark
mass for each configurations.
That covers the strange quark mass range $am_s'=0.0775$-$0.0825$ for $am_s=0.080$ 
and  $am_s'=0.0975$-$0.1025$ for $am_s=0.100$.
To compute the high-mode contributions for several $m_s'$ at once,
we use the multishift solver for the Dirac operator.
We take five random noises for the noisy estimator.
Thus the nucleon mass for each different strange
quark masses is calculated. 
We find small strange quark mass dependence of the nucleon mass 
and its statistical error does not grow for entire region of the
strange quark mass we take.
\begin{figure}[tbp]
  \centering
    \includegraphics[width=6cm,clip]{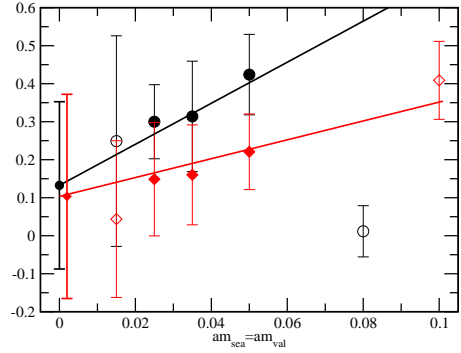}
\caption{
The chiral extrapolation fit for the strange quark content 
for $m_s=0.080$~(black) and $m_s=0.100$~(red).
The plot shows only the data at unitary point. 
The linear line shows the fit functions for unitary point.
For a reference, the empty symbols show the data not used in the fit.
}
\label{fig:mass}
\end{figure}
\section{Extraction of the strange quark content}
To extract the strange quark content $\langle N| \bar{s}s|N\rangle$, 
we fit the nucleon mass by a linear function of strange quark mass.
Then the coefficient of the linear term corresponds 
to the strange quark content of the nucleon.
We use 11 data points for the fit including the unreweighted point at each 
sea and valence quark masses.  
We obtain the non-zero signals of the strange content
at each data point with 50-100\% statistical error. 
The results for the unitary points are shown in Table~\ref{tab:strange}.
\begin{table}[tbp]
  \centering
  \begin{tabular}{ccccccc}
    \hline
$am_{sea}=am_{val}$ & 0.015  & 0.025 & 0.035 & 0.050 & 0.080 & 0.100 \\
    \hline
    $am_s=0.080$   & 0.25(28) & 0.30(10) & 0.31(15) & 0.42(11) & 0.01(7) & -  \\
    $am_s=0.100$   & 0.04(20) & 0.15(15) & 0.16(13) & 0.22(10) & - & 0.41(10)  \\
   \hline
  \end{tabular}
  \caption{
	The strange content of the nucleon at the unitary points.
  } 
  \label{tab:strange}
\end{table}
To estimate the physical values of the strange content, 
we need to carry out a chiral extrapolation.
In order to improve the accuracy, we use partially quenched data 
point.
We take the fit ranges as $am_{sea}=0.025$-$0.050$ and
$am_{val}=0.025$-$0.050$ to avoid large finite volume corrections
and possible higher order corrections.
Its chiral behavior is described by a derivative of 
the nucleon mass formula in SU(3) ChPT in terms of the strange quark mass.
We note that the non-analytic behavior of the light up and down 
quark masses does not appear up to NNLO corrections.
Therefore we safely use the polynomial function 
of the up and down quark masses.
We use a simple fit function 
\begin{eqnarray}
  \label{eq:dmds}
\frac{dM_N}{dm_s}= a_1 + a_2 m_{sea}+ a_3 m_{val}.
\end{eqnarray}
The fit results are shown in Fig.~\ref{fig:mass}.
The data are reasonably fitted by the formula~(\ref{eq:dmds}).
Although our results at chiral limit have about 100\% errors,
it shows that the physical result of the strange quark content is  
very tiny. 
We estimate the renormalization group invariant quantity
$f_{T_s}$ as 
\begin{equation}
  \label{eq:fts}
f_{T_s}=
\left\{
\begin{array}{cr}
0.020(34)\ & (am_s=0.080) \\
0.019(53)\ & (am_s=0.100) \\
\end{array}\right.,
\end{equation}
where the error is statistical.  
There may be some systematic uncertainties coming from the
finite volume corrections and chiral extrapolations.
Both results are consistent each other, 
we find the strange quark mass dependence of this quantity is very
small. 
Using two results of the sigma term obtained in
Sec.~\ref{sec:sigma} and $f_{T_s}$(=0.020), we obtain $y \sim 0.03$.  
These results are comparable to
our previous two-flavor analysis~\cite{Ohki:2008ff}.
We also conclude that the upper bound on $f_{T_s}$
is given by $|f_{T_s}|\le 0.08$ at 1 $\sigma$ level.
\section{Summary}
We calculated the nucleon mass and sigma term 
in 2+1-flavor QCD simulation on the lattice with 
exact chiral symmetry. 
To evaluate the strange quark content of the nucleon, 
we used the reweighting technique for the strange quark mass,
worked very well for studying the 
strange quark mass dependence.
Although our results have large statistical uncertainty,  
we conclude that the strange quark content of nucleon is tiny
compared to its up and down quark contributions.
\vspace{5mm}

The main numerical calculations were performed on IBM System Blue Bene
Solution at High Energy Accelator Organization~(KEK) 
under support of its Large Scale Simulation Program~(No.09-05).
We also used NEC SX-8 at Yukawa Institute for Theoretical
Physics~(YITP), Kyoto University and at Research Center for Nuclear
Physics~(RCNP), Osaka University.
This work is supported in part by the Grant-in-Aid of
the Ministry of Education~(Nos. 21$\cdot$897,
19540286, 
19740160, 
20105001,
20105002,
20105003,
20105005,
20340047,
21674002,
21684013).
The work of HF was supported by the Global COE program of Nagoya
University ``QFPU'' from JSPS and MEXT of Japan.

\end{document}